\documentclass[twocolumn,showpacs,preprintnumbers,amsmath,amssymb]{revtex4}

\usepackage{epsf}
\usepackage{graphicx}
\usepackage{dcolumn}
\usepackage{bm}
\def\be{\begin{equation}}
\def\ee{\end{equation}}
\def\bea{\begin{eqnarray}}
\def\eea{\end{eqnarray}}

\newlength{\dinwidth}
\newlength{\dinmargin}
\begin{document}
 \tighten
\vskip 3cm

\

\title{\Large\bf M-theory, Cosmological Constant and Anthropic Principle   }

\author{\bf Renata Kallosh and Andrei Linde 
}

\affiliation{ {Department
  of Physics, Stanford University, Stanford, CA 94305-4060,
USA}    }

\date{August 26, 2002 \\ \phantom{}}
 
 {
\begin{abstract}We discuss the theory of dark energy based on maximally extended supergravity and suggest a possible anthropic explanation of the present value of the cosmological constant and of the observed  ratio between dark energy and energy of matter.
\end{abstract}}
\pacs{PACS: 98.80.Cq, 11.25.-w, 04.65.+e}

\maketitle


 \section{Introduction: Anthropic constraints on the cosmological constant}

 After  many desperate attempts to prove that the cosmological constant must vanish, now we face an even more complicated problem. We
must understand why the cosmological constant, or the slowly changing dark energy, is at least 120 orders smaller than the Planck density $M_p^4$, and, simultaneously, why its value is as large as $\Omega_D \rho_0$. Here $\rho_0 \sim 10^{-29}$ g/cm$^3 \sim 10^{-120} M_p^4$ is the total density of matter in the universe at present, including the cosmological constant, or dark energy, and $\Omega_D \sim 0.7$. One of the most interesting attempts to provide such an explanation is
related to anthropic principle \cite{barrow}. 

This principle for a long time was rather controversial. It was based on an implicit assumption that the universe was created many times until the final success. It was not clear who did it and why was it necessary to make it  suitable for our existence. Moreover, it would be much simpler to create proper conditions for our existence in a small vicinity of a solar system rather than in the whole universe. 

These problems were resolved with the invention of inflationary cosmology. First of all, inflationary universe itself, without any external intervention, may  produce exponentially large domains with all possible laws of low-energy physics \cite{LindeCam}.  And  if the conditions suitable for our existence are established near the solar system, inflation ensures that similar conditions appear everywhere within the observable part of the universe. 
In addition to considering a single inflationary universe consisting of many domains with different values of constants, one may also consider a possibility that the inflationary  universe may be born in many different quantum states  with different values of coupling constants, see e.g. \cite{Hawking:hk,book,Vilenkin,BellLin}.   This provides a simple justification of the cosmological anthropic principle  and allows one to apply it to the cosmological constant problem.

If, for example, the cosmological constant is large and negative, $\Lambda
\ll -10^{-29}$ g/cm$^3$, then such a universe, even if it is flat, would collapse within the time that is  smaller than the age of our universe $t \sim 14$ billion years
\cite{barrow,linde300}. This would make our life impossible.  One may wonder whether intelligent life could  emerge within $7$ billion years or $5$ billion years, but we have no reason to believe that it could  happen  on a much shorter time scale. 

 \begin{figure}[h!]
\centering\leavevmode\epsfysize=5.2cm \epsfbox{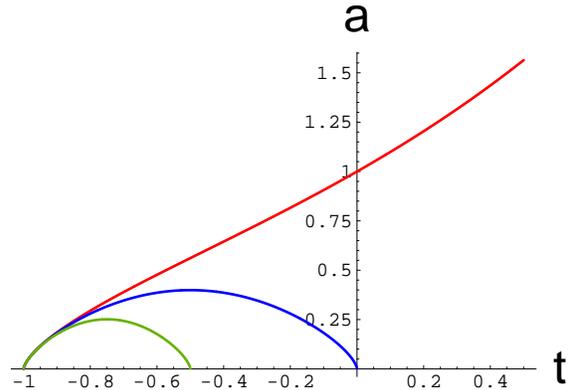}

\

\caption[fig1]
{Evolution of a flat $\Lambda$CDM universe for various values of $\Lambda$. Time is in units of the present age of the universe $t_0 \approx 14$ billion years. The present moment is placed at $t=0$,  the big bang corresponds to $t = -1$. The upper (red) line corresponds to the flat universe with $\Omega_{\rm tot} = 1$,  $\Omega_\Lambda =  0.7$ (i.e. $\Lambda = +0.7 \rho_0$)  and $\Omega_{M} =  0.3$. 
The next line below it (the blue line) corresponds to a flat  universe with $\Lambda = -4.7\rho_0$. As we see, this universe  collapses at the age of 14 billion years. The total lifetime of the universe with $\Lambda = -18.8 \rho_0$ (the lower, green line) is only 7 billion years. }
\label{negative}
\end{figure}

Since we are entering the age of precision cosmology, let us improve the order-of-magnitude estimates of \cite{barrow,linde300} and obtain a  numerical constraint on negative $\Lambda$. The investigation is  straightforward, so we will simply show the results in Fig.  \ref{negative}. We find that the anthropic constraint on the negative cosmological constant  is a bit less stringent than  it was  anticipated.  If $7$ billion years is sufficient  for emergence of human life, then   $\Lambda \gtrsim -18.8 \rho_0\sim - 2 \times 10^{-28}$ g/cm$^3$.   If we really need about $14$ billion years, as one can infer from  \cite{Lineweaver:2002gd}, then the constraint is somewhat stronger: $\Lambda \gtrsim -4.7 \rho_0 \sim -5 \times 10^{-29}$ g/cm$^3$.

However,  the present observational data suggest that $\Lambda >0$. In this case the use of the anthropic considerations become more involved. In \cite{linde300} it was argued that the life of our type is impossible for $\Lambda \gg 10^{-29}$ g/cm$^3$ because in this case the density of matter of the universe would be exponentially small due to its exponential expansion at the present stage. A more precise and rigorous constraint was obtained later by Weinberg \cite{Weinberg87}. He pointed out that the process of galaxy formation occurs only up to the moment when the cosmological constant begins to dominate the energy density of the universe and the universe enters the stage of late-time inflation. (By galaxy formation we understand the growth of density contrast until the moment when the galaxy separates from the general cosmological expansion of the universe. After that, its density rapidly grows by a factor of $O(10^2)$, and its subsequent evolution becomes much less sensitive to the  value of $\Lambda$.) For example, one may consider galaxies  formed  at $z \gtrsim
4$, when the energy density of the universe was 2  orders of
magnitude greater than it is now. Such galaxies would not form if  $\Lambda
\gtrsim 10^2 \rho_0 \sim 10^{-27}$ g/cm$^3$.

Thus, anthropic considerations may reduce the disagreement between the theoretical expectations ($\Lambda \sim M_p^4$) and observational data ($\Lambda \sim \rho_0 \sim 
10^{-120} M_p^4$)  from $120$ orders of magnitude to only 2
orders of magnitude! But this  is not yet a complete solution of the cosmological constant problem. Assuming that all values of the cosmological constant are equally probable, one would find himself in a universe with $\Lambda \sim
\rho_0$ with the probability about $1\%$.  

The next important step  was made in \cite{Efstathiou,Vilenkin:1995nb,Weinberg96,Garriga:1999bf,Bludman:2001iz}. The
authors considered not only our own galaxy, but all other galaxies that
could harbor  life of our type. This would include not only the existing galaxies but also the
galaxies that are being formed at the present epoch. Since the energy density al later stages of the evolution of the universe becomes smaller, even a very small cosmological constant may disrupt the late-time galaxy formation, or may prevent the growth of existing galaxies.   This allowed the authors of \cite{Efstathiou,Vilenkin:1995nb,Weinberg96,Garriga:1999bf,Bludman:2001iz} to strengthen the constraint on the cosmological constant. According to \cite{Weinberg96}, the probability that an astronomer in any of the universes would find a value of $\Omega_D = \Lambda/\rho_0$ as small as $0.7$ ranges from $5\%$ to $12\%$, depending on various assumptions.

However,  our goal is not to find suitable conditions for the human life in general, but rather to explain the results of our observations. These results include the fact that for whatever reason we live in an internal part of the galaxy that probably could not be strongly affected by the existence of a cosmological constant $\Lambda \sim \rho_0$.  Does it mean that we are not typical observers since we live in an atypical part of the universe where we are protected against a small cosmological constant $\Lambda \sim \rho_0$? Also, galaxy formation is not a one-step process. The central part of our galaxy was formed very early, at $z \gtrsim 20$, when the energy density in the universe was 4 orders of magnitude greater than it is now. To prevent formation of such regions one would need to have $\Lambda
\gtrsim 10^4 \rho_0 \sim 10^{-25}$ g/cm$^3$. It may happen that the probability of emergence of life in such regions, or in the early formed dwarf galaxies, is very small. Moreover, one could argue that the probability of emergence of life is proportional to the fraction of matter condensed into large galaxies \cite{Efstathiou,Vilenkin:1995nb,Weinberg96,Garriga:1999bf,Bludman:2001iz}. Even if it is so, in an eternally existing inflationary universe there should be indefinitely many  regions suitable for existence of life, so life would eventually appear in one of such places even if the probability of such event is extremely small. A more detailed investigation of this issue is in order \cite{GarVil}.

In this respect the anthropic constraint on $\Lambda <0$ seems to be  less ambiguous. But it is also less important since it does not seem to apply to an accelerating universe with $\Lambda \approx 0.7 \rho_0$. 
In this paper we will show, however, that a similar constraint based on investigation of the total lifetime of a flat universe can be derived in a broad  class of theories based on N=8 supergravity that can describe the present stage of acceleration \cite{Kallosh:2001gr,Linde:2001ae,Kallosh:2002wj,Kallosh}. This may allow us to avoid fine-tuning that is usually required to explain the observed value of $\Omega_D$.


\section{Maximal Supergravity as Dark Energy Hidden Sector}

Usually in all discussions of the cosmological constant in the astrophysical literature it is assumed that one can simply add a cosmological constant term describing vacuum energy to the gravitational Lagrangian. However, it appears to be extremely difficult to do so in the context of M/string theory.

All known compactifications of the fundamental M/string  theory to four dimensions do not lead to potentials with de Sitter solutions corresponding to $\Lambda > 0$. However, there are versions   of the maximally extended $d=4$ $N=8$ supergravity which have dS solutions.  They are also known to be solutions of $d=11$ supergravity with 32 supersymmetries, corresponding to M/string theory \cite{Hull:1988jw}.  $dS_4$ solutions of $d=4$  $N=8$  supergravity correspond to solutions of M/string theory with non-compact internal  seven (six) dimensional space.   The relation between  states of higher dimensional and four dimensional theory in such backgrounds is complicated since the standard Kaluza-Klein procedure is not valid in this context. It is nevertheless true that the class of $d=4$  supergravities with dS solutions that we will consider below as dark energy candidates has a direct link to M/string theory, as opposite to practically any other model of dark energy.  Moreover, these theories with maximal amount of supersymmetries  are perfectly consistent from the point of view of the $d=4$ theory: all kinetic terms for scalars and vectors are positive definite.

All supersymmetries are spontaneously broken for dS solutions of $N=8$ supergravity. These dS solutions  are unstable; they correspond either to a  maximum of the potential for the scalar fields $\phi$ or to a  saddle point. In all known cases one finds \cite{Kallosh:2001gr} that there is a tachyon and the  ratio between $V''=m^2$ and $\Lambda= V$ at the extremum of $V(\phi)$ is equal to $-2$.

According to current cosmological data, the relevant  $dS_4$ space is defined as a hypersurface in a 5d space
$-T^2+ (X^2+Y^2+Z^2+W^2)=H_0^{-2}  $.
Here $H_0$ is the Hubble parameter. Its inverse,   $H_0^{-1}$, determines time scale of the same order as the age of the universe, $H_0^{-1}\sim 10^{10} \, \rm years $.  
One of the simplest solutions of $d=11$  supergravity is given by a warped product  of a four-dimensional dS space and a seven-dimensional hyperboloid  $ H^{p,q}$. A fiducial model where all scalars are constant is defined by   dS surface presented above and the surface in an eight-dimensional space defining an internal space  hyperboloid $ H^{p,q}$. This surface is given by  $\eta_{AB} z^A z^B= \alpha \, H_0^{-2} $, where the constant $\alpha$ depends on $p,q$ and the metric $\eta_{AB}$ is constant and has $p$ positive eigenvalues and $q$ negative eigenvalues and $p+q=8$.

A simplest (and typical) representative of $d=4$ $N=8$ supergravities with dS maximum, originated from M-theory, has  the following action   \cite{Hull:rt}:

\be\label{pot}
g^{-1/2} L = -{1\over 2} R -  {1\over 2}  (\partial \phi)^2   - \Lambda (2-  \cosh {\sqrt 2} \phi)\ .
 \ee
Here  we use units  $M_{p}=1$. At the critical point
$V'=0 \ , \quad   V_{cr}= \Lambda \ , \quad \phi_{cr}=0
$.

 \begin{figure}[h!]
\centering\leavevmode\epsfysize=5.2cm \epsfbox{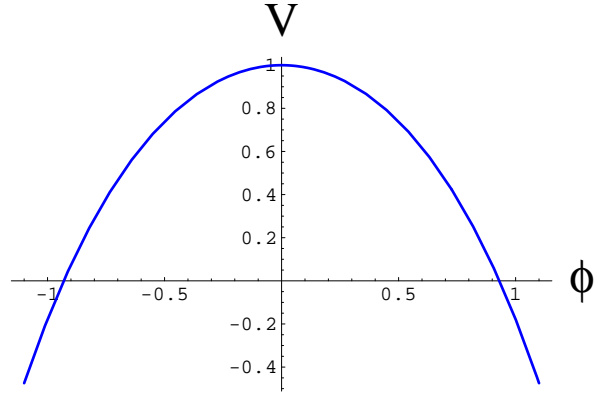}

\

\caption[fig1]
{Scalar potential $V(\phi)=\Lambda (2-  \cosh {\sqrt 2} \phi)$ in $d=4$ $N=8$ supergravity (\ref{pot}). The value of the potential is shown in units of $\Lambda$, the field is given in units of $M_p$.}
\label{Pot}
\end{figure}

This corresponds to the case $p=q=4$ and $d=4$ supergravity has  gauged $SO(4,4)$ non-compact group. At dS vacuum it is broken down to its compact subgroup, $SO(4)\times SO(4)$. The value of the cosmological constant $\Lambda$  is related to $H_0$ and to the gauge coupling $g$ as follows:
$$
\Lambda = 3H_0^2 = 2 g^2 \ .
$$
 
A  similar potential was  obtained  in N=4 gauged supergravity in \cite{Gates:1983ct}, where dS solution of extended supergravity was found for the first time and where it was pointed out that it is unstable, being a maximum of the potential. 

The gauge coupling and the cosmological constant in $d=4$ supergravity have the same origin in M-theory: they come from the flux  of an antisymmetric tensor gauge field strength \cite{Hull:1988jw}. The corresponding  4-form $F_{\mu\nu\lambda\rho}$ in $d=11$ supergravity  is proportional to the volume form of the $dS_4$ space:
\be\label{lam}
F_{0123} \sim  \sqrt {\Lambda} \,  V_{0123} \ .
\ee
Here $F=dA$ where $A$ is a 3-form potential of $d=11$ supergravity. According to this model, the small value of the cosmological constant is due to the  4-form flux which has the inverse  time-scale  of the order of the age of the universe.   Note that in our model there is no reason for the flux quantization  since the internal space is not compact. This makes it  different  from \cite{Bousso:2000xa} where the flux and/or its changes were quantized.  The eleven-dimensional origin of the scalar field $\phi$ in the potential can be explained as follows. Directly in d=4 $N=8$ gauged supergravity has 35 scalars and 35 pseudo-scalars, forming together a coset space ${E_{7(7)}\over SU(8)}$.  The field $\phi$ is an $SO(4)\times SO(4)$  invariant combination of these scalars and it may also be viewed as part of  the $d=11$ metric.

The first idea would be to discard this model altogether because its potential is unbounded from below.  However, the scalar potential in this theory remains positive for $|\phi| \lesssim 1$, and for small $\Lambda$ the time of development of the instability can be much greater than the present age of the universe, which is quite sufficient for our purposes \cite{Kallosh:2001gr,Linde:2001ae,Kallosh:2002wj,Kallosh}. In fact, we will see that this instability allows us to avoid the standard fine-tuning/coincidence problem plaguing most of the versions of the theory of quintessence.
To use these theories to describe the present stage of acceleration (late inflation) one should take $\Lambda \sim   10^{-120}M_{p}^4$. This implies that the tachyonic mass is ultra-light, $|m^2|\sim -(10^{-33}\, \rm{eV})^2$.

In the early universe the ultra-light scalar fields may stay away from the extrema of their potentials; they `sit and wait' and  they begin moving only when the Hubble constant  decreases and becomes comparable with the scale of the scalar mass. This may result in noticeable changes of the effective cosmological constant during the last few billion years. 

Since the potential of $N=8$ supergravity with dS solution is unbounded from below,  the universe will eventually collapse.  If the initial position of the field is not far from the top of the potential, the time before the collapse may be very long \cite{Kallosh}. 

From the perspective of the $d=11$ theory it is natural to consider a large ensemble of possible  values for the fields $F\sim \sqrt{ \Lambda}$ and $\phi$ and study it.  In the context of the $d=4$ theory one may also study a large ensemble of values for $ \Lambda $ and $\phi$.

\section{Anthropic constraints on the cosmological constant in N=8 supergravity}

Consider  a theory of a scalar field $\phi$ with the effective potential $V(\phi)=\Lambda (2-  \cosh {\sqrt 2} \phi)$ in the $N=8$ theory (\ref{pot}). In order to understand the cosmological consequences of this theory, let us first consider this potential at $|\phi| \ll 1$. In this limit the potential has a very simple form,
\be \label{simplepot}
V(\phi) = \Lambda(1 -\phi^2) = 3H^2_0(1 -\phi^2) \ .
\ee
The main property of this potential is that $m^2 = V''(0) = -2\Lambda= -6H_0^2$. One can show that a homogeneous field $\phi \ll 1$ with $m^2 = -6H_0^2$ in the universe with the Hubble constant $H_0$ grows as follows: $\phi(t) = \phi_0 \exp{cH_0t}$, where $c = (\sqrt{33}-3)/2\approx 1.4$. Consequently, in the universe with the energy density dominated by $V(\phi)$ it takes time $t\sim   0.7 H_0^{-1} \ln\phi_0^{-1}$ until the scalar field  rolls down from $\phi_0$ to the region $\phi \gg 1$,  where $V(\phi)$ becomes negative and the universe collapses.  

This means that one cannot take large $\Lambda$ without making the total lifetime of the universe too short  to support life, unless the scalar field $\phi_0$ was exponentially small. But if the potential of the field $\phi$ always was very flat, then one can assume that the field $\phi$ initially (or after inflation) can take any value $\phi_0$ with equal probability, so there is no reason to expect that $\phi_0$ must be very small. This means that for $\phi_0 \lesssim 1$ the typical lifetime of the universe is $t_{\rm tot} \sim H_0^{-1} \sim \Lambda^{-1/2} $. Therefore the universe can live longer than 14 billion years only if the cosmological constant is extremely small, $\Lambda \lesssim \rho_0$.

On the other hand, for $\phi \gg 1$ the potential falls down exponentially, $V(\phi)\sim -\Lambda\, \exp {\sqrt 2   |\phi|}$. Therefore for $\phi \gg 1$ the universe almost instantly collapses even if $\Lambda \lesssim \rho_0$.

Now we can study this process numerically, solving a system of equations for the scale factor $a(t)$ and the scalar field $\phi(t)$.
\bea
&& \ddot \phi + 3{\dot a\over a} \dot \phi - V'(\phi) =0 \ , \nonumber \\
&&  {\ddot a\over a}= {{V -\dot\phi^2-\rho_M }\over 3} \ , \nonumber 
\label{F2}
\eea
where $\rho_M = C/a^3(t)$, and $\phi'_0 = 0$. A detailed description of our approach can be found in \cite{Kallosh}. Here we will present some of our results related to the cosmological constant problem.

\begin{figure}[h!]
\centering\leavevmode\epsfysize=5.2cm \epsfbox{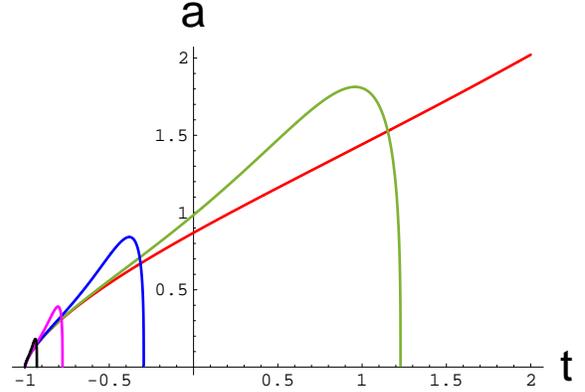}

\

\caption[fig1]
{Expansion of the universe for $\phi_0= 0.25$. Going from right to left, the first (red) line corresponds to $\Lambda = 0.07 \rho_0$, the second (green) line corresponds to $\Lambda = 0.7 \rho_0$, then  $\Lambda =  7\rho_0  \ , 70\rho_0$ and $700\rho_0$. 
}
\label{agelam}
\end{figure}

Fig. \ref{agelam} shows expansion of the universe for $\phi_0= 0.25$, and for various values of $\Lambda$ ranging from $0.7\rho_0$ to $700\rho_0$. Time is given in units of $14$ billion years.  One finds, as expected, that the total lifetime of the universe for a given $\phi_0$ is proportional to $\Lambda^{-1/2}$, which means that large $\Lambda$ are anthropically forbidden.

\begin{figure}[h!]
\centering\leavevmode\epsfysize=5.2cm \epsfbox{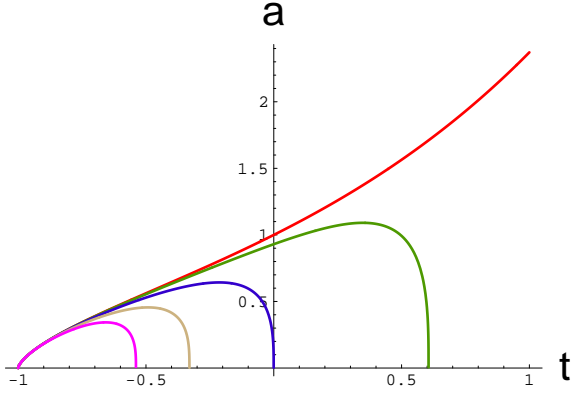}

\

\caption[fig1]
{Expansion of the universe for $\Lambda = 0.7 \rho_0$. The upper (red) line corresponds to the fiducial model with $\phi_0 = 0$ (cosmological constant; field does not move).  The (green) line below corresponds to $\phi=0.5$. The next (blue) line corresponds to $\phi = 1$, then to $\phi = 1.5$, and $\phi = 2$.}
\label{agephi}
\end{figure}

Fig. \ref{agephi} shows expansion of the universe for $\Lambda = 0.7 \rho_0$. The upper (red) line corresponds to the fiducial model with $\phi_0 = 0$. In this case the field does not move and all cosmological consequences are the same as in the standard theory with the cosmological constant $\Lambda = 0.7 \rho_0$.  The difference will appear only in a very distant future, at $t \sim 10^2 H_0^{-1} \sim 10^3$ billion years, when the unstable state $\phi_0 = 0$ will decay due to the destabilizing effect of quantum fluctuations   \cite{Kallosh:2001gr}. For $\phi_0 > 1$ the total lifetime of the universe becomes unacceptably small, which means that large $\phi_0$ are anthropically forbidden.

Further conclusions will depend on various assumptions about the probability of parameters  $(\Lambda,\phi_0)$. In this section we will make  the simplest assumption that all values of $\Lambda$ and $\phi_0$ are equally probable. We will discuss alternative assumptions and their consequences in the next section.

\begin{figure}[h!]
\centering\leavevmode\epsfysize=5.2cm \epsfbox{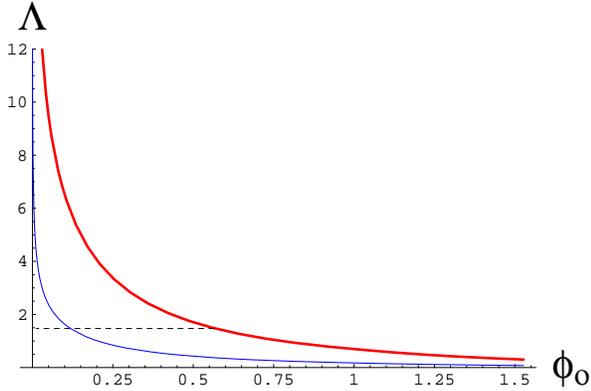}

\

\caption[fig1]
{The region below the thick (red) line contains all possible $\Lambda$ and $\phi_0$ corresponding to the total lifetime of the universe greater than $14$ billion years. The  dashed line $\Lambda \approx 1.5 \rho_0$ separates this region  into two equal area parts.  The region below the thin (blue) curve corresponds to all universes with the lifetime greater than 28 billion years, i.e. to the universes that would live longer than $14$ billion years after the present moment.}
\label{lam2}
\end{figure}

  All possible values of $\Lambda$ and $\phi_0$ corresponding to the total lifetime of the universe greater than $14$ billion years are shown in Fig. \ref{lam2}, in the region under the thick (red) line. If all values of   $\Lambda$ and $\phi_0$ are equally probable, the measure of probability is given by the area. The total area under the curve is finite, $S_{\rm tot} \approx 3.5$. One can easily estimate the probability to be in any region of the phase space $(\Lambda,\phi_0)$ by measuring the corresponding area and dividing it by  $S_{\rm tot}$.

The  dashed line $\Lambda \approx 1.5 \rho_0$ separates the anthropically allowed region  into two equal area parts.  This implies that  the average value of $\Lambda$  in this theory is about $1.5\rho_0$. It is obvious that $\Lambda$ can be somewhat larger or somewhat smaller than $1.5\rho_0$, but the main part of the anthropically allowed area corresponds to 
$$\Lambda = O(\rho_0) \sim  10^{-120} M_p^4 .$$  
This is one of the main results of our investigation. This result is a direct consequence of the relation $m^2 = -6H_0^2$ which is valid for all known versions of $d=4$ $N=8$ supergravity that allow dS solutions.

The region below the thin (blue) curve corresponds to all universes with the lifetime greater than 28 billion years, i.e. to the universes that would live longer than $14$ billion years after the present moment. The area below this curve is 3 times smaller than the area between the thin (blue) curve and the thick (red) curve. This means that the ``life expectancy'' of a typical anthropically allowed universe (the time from the present moment until the global collapse)  is smaller than the present age of the universe. The prognosis becomes a bit more optimistic if one takes into account that we live in the universe with $\Omega_D = 0.7$: The probability that the universe will survive more than 14 billion years from now becomes better than 50\%.

\begin{figure}[h!]
\centering\leavevmode\epsfysize=5.9cm \epsfbox{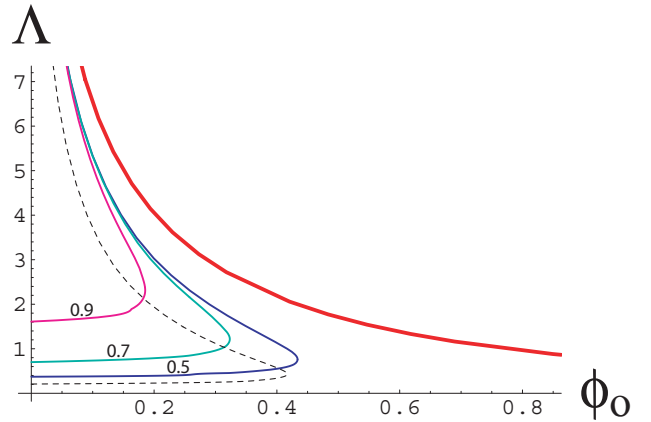}

\

\caption[fig1]
{The curves corresponding to $\Omega_D = 0.5$ (blue line below the thick red line), $\Omega_D = 0.7$ (green line), and $\Omega_D = 0.9$ (pink  line). The region to the right of the thin dashed line corresponds to the universes that accelerate  14 billion years after the big bang.}
\label{combo}
\end{figure}

Finding the average value of $\Lambda$ does not immediately tell us   what is the most probable value of $\Omega_D$.
In order to do it we plot in Fig. \ref{combo} the curves corresponding to $\Omega_D = 0.5$ (blue line, just below the thick red line), $\Omega_D = 0.7$ (green line, just below the line $\Omega_D = 0.5$), and $\Omega_D = 0.9$ (pink  line, below the line $\Omega_D = 0.7$). The region to the left of the thin dashed line corresponds to the universes that accelerate at the present time (14 billion years after the big bang).

 The total probability to live in an accelerating universe at the time 14 billion years after the big bang is determined by the area bounded by the thin dashed line in Fig. \ref{combo}. Dividing this area by $S_{\rm tot} \approx 3.5$, one can find that this probability is about 35\%. About a half of this area corresponds to $\Omega_d > 0.9$.
The most interesting part of the accelerating region is bounded by the  blue  curve $\Omega_D = 0.5$, the pink curve $\Omega_D = 0.9$ and the thin dashed line. All points inside this region correspond to accelerating universes with $0.5 < \Omega_D < 0.9$. As one can easily see from Fig. \ref{combo}, the area of this region is about $0.4$. Dividing it by the total area of the anthropically allowed region $S_{\rm tot} \approx 3.5$ one finds that the probability to live in such a universe is about 10\%. These results  resolve  the fine-tuning/coincidence problem in the theory of dark energy.

\section{Discussion}
Most of the theories of dark energy have to face two problems. First of all, it is necessary to explain why does the bare cosmological constant vanish. Then one must find a dynamical mechanism imitating a small cosmological constant and explain why $\Omega_D\sim 0.7$ at the present stage of the evolution of the universe.

In this paper we studied cosmological consequences of the simplest toy model of dark energy based on $N=8$ supergravity. We have found that in the context of this theory one can completely resolve the cosmological constant problem, as well as the coincidence problem plaguing most models of quintessence.

Indeed, one simply cannot add a cosmological constant to this theory.   The only way to introduce something similar to the cosmological constant is to put the system close to the top of the effective potential. If the potential is very high, then  it is also very curved, $V''(0) = -2V(0)$. We have found that the universe can live long enough only if  the field $\phi$ initially is within the Planck distance from the top, $|\phi|\lesssim M_p$, which sounds reasonable, and if   $V(0)$, which plays the role of $\Lambda$ in this theory, does not exceed much the critical value $\rho_0 \sim 10^{-120} M_p^4$.

In our paper we made the simplest assumption that the probability to live in the universe with different $\Lambda$ and $\phi_0$ does not depend on their values. However, in realistic models the situation may be different. For example, as we mentioned,  $\Lambda^{1/2}$  is related to the  4-form flux in $d=11$ supergravity, see Eq. (\ref{lam}). This may suggest that the probability distribution should be uniform not with respect to $\Lambda$ and $\phi_0$ but with respect to $\Lambda^{1/2}$ and $\phi_0$. We studied this possibility and found that the numerical results change, but the qualitative features of the model remain the same.

The probability distribution for $\phi_0$ also may depend on $\phi_0$ even if $V(\phi)$ is very flat at $\phi <1$. First of all, it might happen that the fields $\phi \gg 1$ (i.e. $\phi \gg M_p$) are  forbidden, or the effective potential at large $\phi$ blows up. This is often the case in $N=1$ supergravity. Secondly, interactions with other fields in the early universe may create a deep minimum capturing the field $\phi$ at some time-dependent point $\phi < 1$. This also often happens in $N=1$ supergravity, which constitutes one of the features related to the cosmological moduli problem. If this happens in our model, one will be able to ignore the region of $\phi_0>1$ (the right part of Figs. \ref{lam2}, \ref{combo}) in the calculation of probabilities. This will increase the probability to live in an accelerating universe  with $0.5 < \Omega_D < 0.9$.

In our estimates we assumed that the universe must live as long as 14 billion years before the human life appears. One could argue that the first stars and planets were formed long ago, so we may not need much more than about 5-7 billion years for the development of life. This would somewhat decrease our estimate for the probability to live in an accelerating universe  with $0.5 < \Omega_D < 0.9$, but this would not alter our results qualitatively. On the other hand, most of the planets were probably formed at later stages of the evolution of the universe \cite{Lineweaver:2002gd}, so one may argue that the probability of emergence of human life becomes much greater at $t > 14$ billion years, especially if one keeps in mind how many coincidences  have made our life possible. If one assumes that human life is extremely improbable (after all, we do not have any indications of its existence anywhere else in the universe), then one may argue that the probability of emergence of life becomes significant only if the total lifetime of the universe can be much greater than 14  billion years. This would increase our estimate for the probability to live in an accelerating universe  with $0.5 < \Omega_D < 0.9$. 

So far we did not use any considerations based on the theory of galaxy formation \cite{Weinberg87,Efstathiou,Vilenkin:1995nb,Weinberg96,Garriga:1999bf,Bludman:2001iz}. If we do so,  the  probability of  emergence of life for $\Lambda \gg \rho_0$ will be additionally suppressed, which will  increase the probability to live in an accelerating universe  with $0.5 < \Omega_D < 0.9$.

To the best of our knowledge, only in the models based on extended supergravity the relation $|m^2| \sim H^2$, together with the absence of freedom to add the bare cosmological constant,  is a property of the theory rather than of a particular dynamical regime. That is why the increase of $V(\phi)$ in such models entails the increase in $|m^2|$. This, in its turn, speeds up the development of the cosmological instability, which  leads  to the anthropically unacceptable consequences.

The $N=8$ model discussed in our paper is just a toy model. In this model we were able to find a complete solution to the cosmological constant problem and to the coincidence problem (explaining why $\Lambda \sim \rho_0$ and why $\Omega_D$ noticeably differs both from $0$ and from $1$ at the present stage of the evolution of the universe). This model has important advantages over many other theories of dark energy, but
to make it fully realistic one would need to construct a complete theory of all fundamental interactions including the dark energy sector described above. This is a very complicated task that is beyond the scope of the present investigation.  
However, most of our results are not model-specific. For example, instead of $N=8$ supergravity one could study any model with the effective potential of the type $V(\phi) = \Lambda(1 -\alpha\phi^2)$ with $\alpha = O(1)$. Another example is provided by the simplest N=1 Pol\'{o}nyi-type SUGRA  model with a very low scale of supersymmetry breaking and with a minimum of the effective potential at $V(\phi) <0$. As shown in \cite{Kallosh}, models of this type also have the crucial property $|m^2| \sim H^2$. In fact, this property is required in most of the models of quintessence.  

Therefore it would be interesting to apply our methods to the models not necessarily related to extended supergravity. A particularly interesting example is the axion quintessence model. The original model suggested in \cite{Frieman:1995pm} has the  potential  
$\Lambda [\cos \left(\phi/f\right)+C]$, and it was assumed  that $C=1$. It was however emphasized in \cite{Frieman:1995pm} that this is just an assumption. The positive definiteness of the potential  with $C=1$ and the fact that it has a minimum at $V=0$ could be motivated, in particular, by the global supersymmetry arguments. In supergravity and M/string theory these arguments are no longer  valid and the derivation  of the value of the  parameter $C$ is not available. In \cite{Choi:1999xn} the axion model of quintessence was studied using the arguments based on M/string theory.  The potential was given in the form
$
V= \Lambda \cos \left(\phi/f\right)
$ without any constant part. 
 
This  potential has a maximum at $\phi = 0$, $V(0)=\Lambda$. The universe collapses when the field $\phi$  rolls to the minimum of its potential  $V(f\pi)=-\Lambda$.
The curvature of the effective potential in its maximum is given by
$
m^2= -{\Lambda/f^2} = -{3H_0^2/f^2} $.
For $f = M_p = 1$  one finds
$
m^2= -{\Lambda} = -{3} H_0^2,
$
and for $f = M_p/\sqrt 2$ one has
$
m^2= -{2\Lambda} = -{6} H_0^2$,
exactly as in the $N=8$ supergravity. Therefore  the anthropic constraints on $\Lambda$ based on the investigation of the collapse of the universe in this model (for $C=0$) are similar to the constraints obtained in our paper for the $N=8$ theory \cite{KKL}.  
However, in this model, unlike in the models based on extended supergravity, one can easily add or subtract any value of the cosmological constant. In order to obtain 
useful anthropic constraints on the cosmological constant in this model one should use a combination of our approach with the usual approach based on the theory of galaxy formation \cite{Weinberg87,Efstathiou,Vilenkin:1995nb,Weinberg96,Garriga:1999bf,Bludman:2001iz}.

In this sense, our main goal was not to replace the usual anthropic approach to the cosmological constant problem, but to suggest its possible enhancement. We find it very encouraging that  our approach may strengthen the existing anthropic constraints on the cosmological constant in the context of the theories based on extended supergravity.

One may find it hard to believe that in order to explain the results of cosmological observations one should consider theories with an unstable vacuum state. However, one should remember that exponential expansion of the universe during inflation, as well as the process of galaxy formation, is the result of the gravitational instability, so we should learn how to live with the idea that our world can be unstable. Also, we did not willingly pick up the theories with an unstable vacuum. We wanted to find the models based on M/string theory that would be capable of describing de Sitter state. All models related to M/string theory that we were able to find so far, with exception of the recently constructed model based on $N=2$ supergravity \cite{Fre:2002pd}, lead to unstable dS vacuum. So maybe we need to take this instability seriously.

This brings us good news and bad news. The bad news  is that in all the theories we have considered in this paper, our part of the universe is going to collapse within the next 10-20 billion years or so.

The good news is that we still have a lot of time to find out whether this is really going to happen.

 \subsection*{ Acknowledgements}
It is a pleasure to thank  T. Banks, M. Dine, T. Dent, J. Frieman, N. Kaloper, A. Klypin,  L. Kofman, D. Lyth,  L. Susskind, A. Vilenkin and S. Weinberg for useful discussions. This work
was supported by NSF grant PHY-9870115. The work by A.L. was also supported
by the Templeton Foundation grant
No. 938-COS273.

\end{document}